\documentclass[fleqn,11pt]{article}

\usepackage[utf8]{inputenc}
\usepackage[T1]{fontenc}
\usepackage{mathtools} 
\usepackage[mathlines]{lineno} 
\usepackage{soul} 
\usepackage{upgreek}
\usepackage{amsmath}
\usepackage{amssymb}
\usepackage{bm}
\usepackage{siunitx}
\usepackage{color}
\usepackage{tabularx}
\usepackage{graphicx}
\usepackage{xspace}
\usepackage{float}
\restylefloat{table}
\usepackage[font={sf,small},labelfont=bf]{caption}
\usepackage{xurl}
\usepackage[numbers,super,sort&compress]{natbib}
\usepackage[colorlinks,citecolor=blue,linkcolor=blue]{hyperref}
\usepackage[normalem]{ulem}
\usepackage[misc]{ifsym}

\title{\textbf{\textsf{Ultra-transient grating spectroscopy for visualization of surface acoustics}}}

\date{}

\author{Tom\'{a}\v{s} Grabec$^1$, Pavla Stoklasov\'{a}$^{1}$, Krist\'{y}na Rep\v{c}ek$^{1}$, Jakub Ku\v{s}n\'{i}r$^{1,2}$, David Mare\v{s}$^{1}$,\\
Martin \v{S}ev\v{c}\'{i}k$^{1}$, Petr Sedl\'{a}k$^{1}$, Hanu\v{s} Seiner$^{1,\mbox{\footnotesize{ \Letter}}}$}

\begin{document}

\maketitle

\noindent

\medskip

\noindent
$^1${\it Institute of Thermomechanics, Czech Academy of Sciences, Dolej\v{s}kova 1402/5, 182 00 Prague, Czechia}

\medskip
\noindent
$^2${\it Faculty of Nuclear Sciences and Physical Engineering, Czech Technical University in Prague, Trojanova 13, 120 00 Prague, Czechia}

\bigskip
\noindent
$^{\mbox{\footnotesize{ \Letter}}}$ {{Corresponding author:} hseiner@it.cas.cz (H. Seiner)}

\bigskip

\begin{abstract}

\textbf{
    \small
    	 Ultrasonic wave propagation across material surfaces reveals essential information about the materials' elastic behavior. All direction- and frequency-dependent features of the propagation are encoded in the Green’s function. Here we present a direct experimental visualization of the frequency-domain angle-resolved Green’s function, including all its complexity resulting from elastic anisotropy. We achieve this visualization using transient grating spectroscopy (TGS), otherwise well established for measuring Rayleigh-type surface acoustic waves (SAWs). But here we focus on early-time thermoacoustic phenomena in the TGS experiment, revealing that, along with the transient standing-wave patterns of SAWs, there also emerge oscillations with at least an order of magnitude shorter lifetimes. These oscillations superpose into dynamic displacement patterns that are transient with respect to the classical transient timescales in TGS; the optical diffraction signal from these ultra-transient gratings enables capturing the surface acoustic response in full detail. By utilizing this feature, ultra-transient grating spectroscopy (UTGS) opens pathways for studying anisotropic materials.}
\end{abstract}

\section*{Introduction}
From the crystal lattice at the atomic scale to the mesoscopic layout of artificially built composites, the internal structure of materials manifests itself in their dynamic mechanical response.
This includes direction-dependent (anisotropic) properties and frequency-dependent (dispersive) behavior.\cite{every_bulk_2013}
Capturing the full complexity of the dynamic response therefore requires a description in the frequency--direction parameter space.
For acoustic experiments on free surfaces of opaque materials, this implies a frequency-resolved description of the material for all directions lying along the surface.\cite{Every2002}

The theoretical object representing the response of the surface is the frequency-domain surface elastodynamic Green's function, which is a function describing the dynamic displacement field arising on a surface of a half-space when loaded by a force that is harmonic both in time and space -- a force with prescribed surface wavevector ($k$-vector) and frequency.\cite{Every1997}
The frequency-domain Green's function is difficult to obtain experimentally because it requires detecting the displacement field immediately after the force is applied and at the same location.
However, the state-of-the-art laser-ultrasonic techniques for materials characterization are not far from accomplishing this task.
Spatially harmonic sources are routinely used in the transient grating spectroscopy method (TGS, setup shown in Figure \ref{fig1}(a)), which is a technique utilizing two coherent laser beams to create interference fringes on the surface of the sample (Figure \ref{fig1}(b)).\cite{rogers1997,Maznev1998,hofmann_transient_2019} 
The interference pattern acts as an array of line-like thermoacoustic sources (Figure \ref{fig1}(c)) with a harmonic energy  perpendicular to the fringes; this direction determines the orientation of the surface $k$-vector of the acoustic waves generated by the pattern, and can be changed by rotating the sample with respect to the optical path.\cite{Stoklasova2021}
In the time domain, the TGS source is a short pulse, spanning a broad frequency band from which any chosen frequency can be analyzed separately using spectral methods. 
This means that the TGS technique is, in theory, perfectly suitable for fulfilling the assumptions for the frequency-domain Green's function in terms of the spatio-temporal character of the prescribed force. 
But is it also capable of detecting the immediate response of the surface to this force?

In the traditional interpretation of the TGS experiment on ideally opaque samples,
	the interference pattern, shown in Figure \ref{fig1}(e), generates an oscillating standing-wave field forming a ripple of the surface --- the transient grating --- created by counter-propagating planar surface acoustic waves (SAWs).
	The standing-wave field arises as a superposition of pulse-like SAWs propagating from individual sources (Figure \ref{fig1}(f)); as the amplitude of SAWs does not significantly diminish with propagation, the standing wave they form can be understood as a local transient resonance of the free surface, which is also where the energy remains localized.
	These vibrations are then optically detected by the diffraction of the probe beam, and the spectrum of the recorded diffraction signal is typically understood as the main acoustic output of the TGS experiment.\cite{maznev_surface_1999,Stoklasova2021}
	The long-duration oscillation of SAWs leads to the formation of dominant sharp peaks in the frequency spectra, at the frequency given by the velocity of SAWs and the wavelength determined by the spatial periodicity of the source.
	 Similar standing wave patterns, easily detectable by TGS, are also created by pseudo-Rayleigh waves (pSAWs).\cite{Farnell1970}
Besides the oscillatory part of the signal arising from the acoustic standing waves, the ripple also has a non-acoustic part formed due to thermal expansion (the transient thermal grating), that can be utilized for measuring the thermal diffusivity of the material, which is also broadly utilized in TGS experiments.\cite{Johnson2012,Huberman2017,Dennett2018}

However, as we already noticed in previous works,\cite{Stoklasova2021, ZoubkovaIEEE2021,song_lightweight_2025,repcek_compliant_2024,Grabec2024_NiTi} the TGS experiments have a certain ability to detect acoustic features beyond the SAW and pSAW standing waves. An edge-like high-frequency contrast in angular maps corresponding to longitudinal wave velocity was noted in Ref.\citenum{Stoklasova2021}, and the whole complexity of the experimental maps in Ref.\citenum{ZoubkovaIEEE2021}.
	Even without understanding the underlying mechanism, the presence of weak, bulk wave-related features can be utilized as additional information about the elastic properties of the examined material.
	We have recently demonstrated this on examples of superelastic\cite{song_lightweight_2025} and magnetic shape memory\cite{repcek_compliant_2024} single crystals, as well as thin epitaxially grown films.\cite{Grabec2024_NiTi}
	In parallel to our work, weak features corresponding to longitudinal bulk wave velocity were found in TGS measurements of a mantis-shrimp club,\cite{alderete_science_2025}
	and some very fine, low-amplitude features at frequencies above the dominant SAW peak can be found in the spectra shown in recent publications,\cite{trachanas2023,Rajagopal2025} although without further analysis or comment. All the above experimental observations motivated us to investigate the origin of these  phenomena.

	In this work, we show that all these additional features can be rationalized by analyzing the early-time response of the surface to the spatially harmonic thermoacoustic source used in TGS. We demonstrate that for both weakly and strongly anisotropic single-crystalline materials, and for both principal and general crystallographic orientations of the free surface, the response of the material recorded in the TGS experiment is, indeed, a close approximation of the respective component of the frequency-domain Green's function. In other words, we show that the TGS method, with slight but targeted modifications, allows detailed visualization of this component. This enables us to study several fine features in the surface acoustics, to explain these features using the framework of homogeneous and evanescent partial waves\cite{Every1997}, and to discuss how these features can be utilized in materials characterization using laser ultrasonics.

\section*{Results}
\subsection*{Introducing the Concept of Ultra-Transient Gratings}

\begin{figure}[!t]
	\centering
		\includegraphics[width=\textwidth]{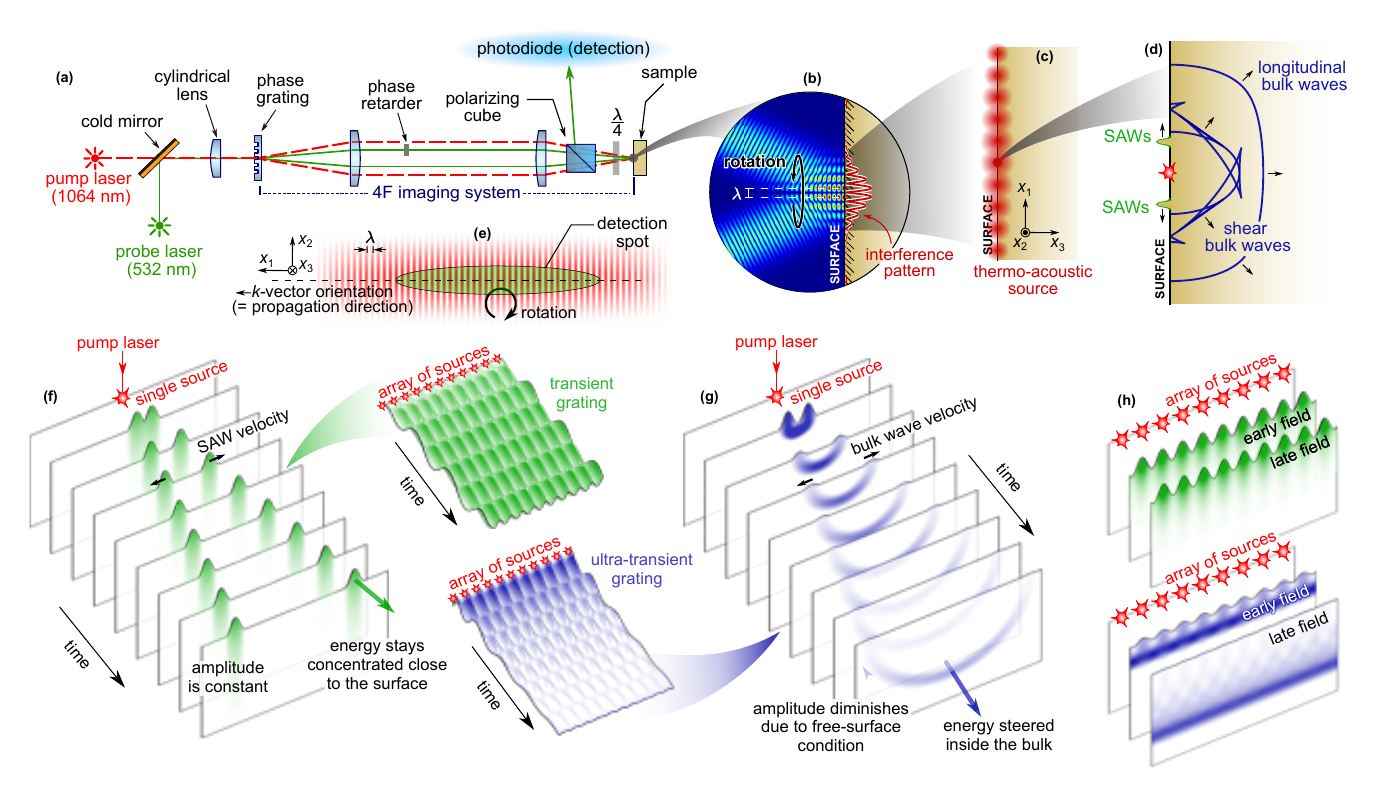}
	\caption{
		{\bfseries TGS experimental setup and visualization of transient and ultra-transient gratings.  }\\
		{\bfseries (a)} the optical path used for TGS measurements, see {\color{blue} Supplementary Figure S1} for a physical realization;
		{\bfseries (b,c)} formation of a spatially harmonic thermoacoustic source in the TGS experiment, the source arises from an interferometric pump-laser pattern with spatial period $\lambda$, and the rotation of the sample enables measurements in different directions along the surface; 
		{\bfseries (d)} a hypothetical response of the anisotropic free surface to one line-like source from the TGS source pattern; 
		{\bfseries (e)} the geometry of the pump laser interference pattern, and the shape and relative size of the detection spot;
		{\bfseries (f,g)} temporal and spatial evolution of the displacement field connected to SAW and bulk wave, respectively, with the excitation by a single source (line), with the adjacent figure showing the displacement at the surface originating from an array of sources;
		{\bfseries (h)} comparing the early and late displacement fields for surface and bulk acoustic waves also below the surface, illustrating the difference between transient grating formed by SAWs and ultra-transient grating formed by bulk waves.
	}\label{fig1}
	\end{figure}

Looking more closely at the thermoacoustic source presented by each individual fringe in the TGS experiment, it generates not only surface acoustic waves but also bulk acoustic waves, that propagate from the source into the whole half-space (Figure \ref{fig1}(d)).\cite{royer2001}
Unlike for the SAWs or pSAWs, the amplitude of the bulk waves therefore diminishes with the propagation, as dictated by the increasing area of the wavefront.
At the same time, the stress field carried by the bulk wave often violates the free-surface (plane stress) conditions, which leads to further suppression of their amplitude close to the surface.
Because of both these effects, the amplitude of the out-of-plane displacements corresponding to the bulk wave propagation at the surface quickly diminishes, and its energy is steered into the bulk, as visualized in Figure \ref{fig1}(g).
Still, superposition of these displacement fields for a periodic array of sources can lead to a formation of a standing-wave pattern -- a very short-lived grating, which we henceforth call the ultra-transient grating, to emphasize that this grating is transient with respect to the transient grating created by standing SAWs.
Since the TGS signals are recorded continuously across the loading pulse moment and directly at the location where the pulse is applied, they should also contain information about the early-time, ultra-transient response -- albeit small in amplitude and short in duration, and therefore small in frequency spectra.
We will further distinguish between the early-time fields, where both the transient and the ultra-transient gratings contribute to the diffraction of the probe beam, and late-time fields, where only the transient gratings do (Figure \ref{fig1}(h)).
	
In the following, we demonstrate that the TGS experiment can, due to detecting the ultra-transient phenomena along with the standard transient gratings, lead to experimental reconstruction of the frequency-domain Green's function with remarkable accuracy.
For this capability of the TGS experiment, that is, for capturing the fine features of the Green's function due to the presence of ultra-transient gratings, we will introduce the term ultra-transient grating spectroscopy (UTGS).
The UTGS approach can offer detailed experimental information on functional materials and bring direct insight into the complex elastodynamic behavior of their free surfaces
	-- behavior that, while theoretically well explored, is challenging to obtain experimentally.
 To demonstrate this, we perform experiments on various cuts of single-crystalline samples, where the complexity of the response follows from elastic anisotropy, and for which the theoretical {frequency-domain} Green's function can be easily calculated, provided that the elastic constants are known, thus enabling a straightforward comparison between theory and experiment.
Nevertheless, the experimental approach is suitable not only for single-crystalline materials, but also for surfaces of heterostructures (such as polycrystals, layered media, and composites) or acoustic metamaterials,\cite{rogers2000,VegaFlick2017,Grabec2024_NiTi} for which the calculation of the Green's function can be much less straightforward.

	\subsection*{Enhancing the Ultra-Transient Gratings  in TGS Experiments}

	The manifestations of the ultra-transient mechanism in the TGS spectra in prior works\cite{Stoklasova2021, ZoubkovaIEEE2021,song_lightweight_2025,repcek_compliant_2024,Grabec2024_NiTi,alderete_science_2025,trachanas2023,Rajagopal2025} were either ignored or utilized only to a minimal extent, and without understanding their origin. With recognizing this mechanism, however, the TGS technique gains broader functionality, significantly extending the information that can be obtained about the examined material from the measurement. This motivated us to identify which modifications make the TGS more sensitive to {the ultra-transient} phenomena, and we developed them further.

	First, we optimized the optical geometry:
	We utilize strongly elliptical beam shapes\cite{maznev_surface_1999} (long-to-short axis ratio of 8 to 1, see Figure \ref{fig1}(e)) for the pump laser, to maximize the number of fringes (periods) with strong excitation.
	To maximize the diffraction potential, we shape the detection beam to a similar elliptical cross-section, with the axes lengths just above half that of the excitation.\cite{Siegman1977}
	That ensures that we obtain the information from locations of maximal excitation (with most of the pump-laser power).
	Also, we strengthen the diffraction efficiency by using two distinct laser wavelengths (\SI{1064}{\nano\metre} for pump and \SI{532}{\nano\metre} for probe), which allows us to leave them in the same sagittal plane, enhancing both sensitivity and $k$-vector selectivity of the measurement.\cite{maznev_surface_1999}
		This choice brings other problems -- mainly the loss of diffraction efficiency on the phase mask and vulnerability to chromatic aberration due to the large difference of optical wavelengths used.
		The former can be accepted -- the phase mask is optimized for the probe-beam wavelength (\SI{532}{\nano\metre}), and the pump laser has a power reserve (and we tune the power output by neutral-density filter at the beginning of the pump-beam path to stay below the ablation limit).
		The latter (chromatic aberration) is solved by the use of achromatic lenses and by the 4F system itself, which compensates the aberrations.

	Second, we utilized the full potential of the probe-beam optical signal with highly sensitive photodiodes and a specialized amplification system.
		Specifically, we use an in-house designed combination of Si PIN photodiodes with \SI{60}{\decibel} amplifiers, thus combining high sensitivity, maximal power input, and amplification to significantly increase the signal-to-noise ratio. 
		The photodiodes also allow for higher maximal power of the probe beam, which translates into better signal-to-noise ratio (following the equations for intensity of heterodyned signal \cite{maznev_surface_1999,Verstraeten2015}).
		The frequency bandwidth of both photodiodes and amplifiers is \SI{1}{\giga\hertz}.
		To fully utilize this bandwidth, we set the geometry to the acoustic wavelength of $\lambda =$\SI{10}{\micro\metre}.
		It is common for TGS experiments to use lower acoustic wavelength 
		-- typically for characterization of thin films,\cite{maznev_surface_1999,rogers2000,Sermeus2014} surface damage from irradiation,\cite{Hofmann2015,Duncan2016,trachanas2023} or for spatial modulus mapping,\cite{reza2020,weaver2024}
		-- where shorter wavelengths (typically down to 2-\SI{6}{\micro\metre}) lead to better localization.
		However, for such wavelengths, the common \SI{1}{\giga\hertz} bandwidth is well-suited for capturing the dominant frequency peak corresponding to SAW/pSAW -- but not that of the faster acoustic waves in typical single crystalline solids (metals, inorganic oxides).
		Note also that the longer wavelength makes the pulsed laser source (with a typical pulse duration  of $\sim\SI{0.5}{\nano\second}$) shorter with respect to the characteristic times in the wave field, 
			leading to effective broadband excitation of all acoustic modes.

	 Third, since the ultra-transient features are expected to have small amplitudes, we maximized the signal-to-noise ratio through extensive averaging -- up to 50 thousand shots per direction -- combined with a fine angular step, which makes otherwise weak features traceable through their continuous directional evolution.
		This results in a scanning time for an angular-dispersion map of approximately 5 hours.
		Due to the collinear geometry of pump and probe beams, the measurement is stable over such timescales without any necessary adjustments, and the results are fully reproducible.

		Further details on all three points above — including a pulse-duration analysis and comparisons at lower averaging and coarser angular steps — are given in the {\color{blue}{Supplementary Discussion}}.

	\subsection*{Comparing the Measured and the Calculated}
	
\begin{figure}[!t]
	    \centering
	    \includegraphics[width=\textwidth]{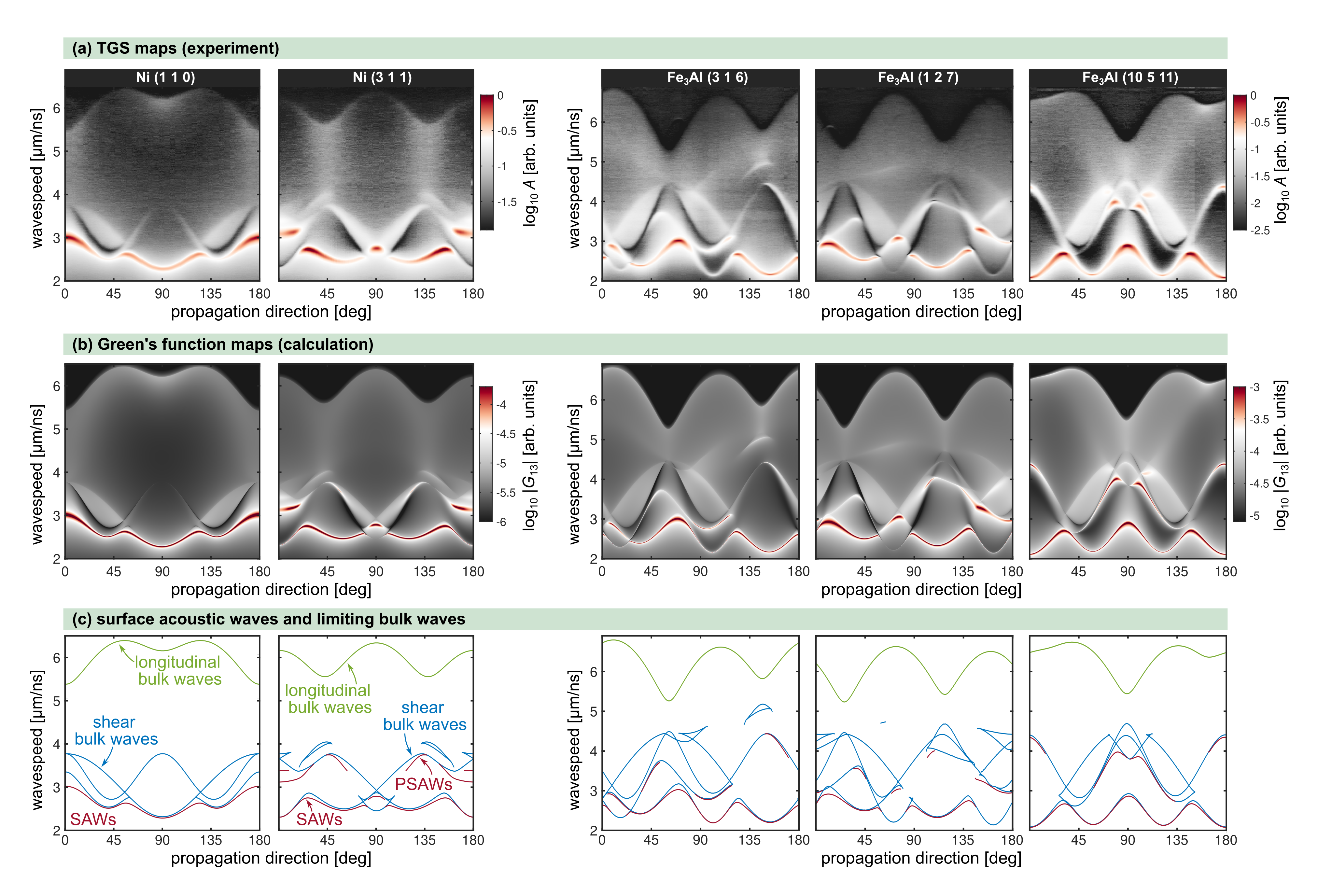}
	    \caption{
	    	{\bfseries Measured and calculated frequency-domain angle-resolved maps on cuts of anisotropic single crystals. }
	    	{\bfseries (a)} experimental maps for high-symmetry planes of Ni (the left  two columns) and low-symmetry planes of {Fe\textsubscript{3}Al}\ (the right three columns);
	    	{\bfseries (b)} corresponding maps of the $|G_{13}|$ Green's tensor component; 
	    	{\bfseries (c)} corresponding wavespeeds of surface-guided and bulk acoustic waves. The bulk waves are the so-called limiting bulk waves, which means bulk waves that carry energy flux in direction parallel to the free surface (several such waves can exist for the shear modes for each direction of propagation).
	    	}
	    	\label{fig2}
	     \end{figure}

	With these modifications, the TGS method has been applied to a nickel single crystal and an iron aluminide single crystal ({Fe\textsubscript{3}Al}).
	For each free surface, we performed a 0--\ang{180} angular scan with a \ang{1} step, with {the resulting frequency-domain TGS maps} shown in Figure \ref{fig2}(a).
	To avoid explicit dependence on the used wavelength, the maps are shown in the direction-vs-wavespeed coordinates, where the wavespeed $v$ was recalculated from the frequency $f$ and wavelength $\lambda$ as $v=\lambda{}f$.
	These two materials were chosen as prototypes of cubic single crystals with moderately strong elastic anisotropy (Zener anisotropy ratio\cite{zener_elasticity_1948} for Ni is $Z=2.6$) and strong elastic anisotropy ($Z=4.3$ for {Fe\textsubscript{3}Al}), respectively.
	Also, for both materials, the wavespeeds of acoustic waves (all modes of bulk acoustic waves in all directions, as well as SAWs/pSAWs on arbitrary cuts) fall between 2 and \SI{7}{\micro\metre\per\nano\second}, which corresponds to the frequency range of 200-\SI{700}{\mega\hertz}, covering perfectly most of the bandwidth available.
	On the Ni single crystals, the measurements were performed on high-symmetry cuts by (1\,1\,0) and (3\,1\,1) lattice planes, while on general irrational planes for {Fe\textsubscript{3}Al}\ -- the approximate Miller indices of the cuts are given in the headings to individual maps in Figure \ref{fig2}(a), precise orientation is specified in {\color{blue}{Supplementary Table S1}}.

	The results show that although both materials have the same crystal symmetry, the obtained maps increase in complexity with increased strength of elastic anisotropy as well as with reduced symmetry of the cuts:
	All the maps shown, and particularly those for the general cuts of the strongly anisotropic {Fe\textsubscript{3}Al}\ crystal, exhibit a variety of peaks, frequency bands, shoulders, and discontinuities, making it obvious that they include much richer information than just frequencies of standing SAW/pSAW wave patterns for given directions.
	Compared to previously published TGS maps\cite{Stoklasova2021,repcek_compliant_2024}, all fine features in Figure \ref{fig2}(a) are considerably sharper and more distinct --- we attribute this to the modifications of the TGS experiment and signal processing discussed in the previous section and in Methods.
	But how are these features related to the respective {frequency-domain} Green's functions?
	
	\medskip
	
	For the opaque metallic samples studied here, the optical penetration depth at the pump wavelength is of the order of tens of nanometres (see Methods), confining the thermoacoustic source to a surface layer of negligible thickness compared to the \SI{10}{\micro\metre} grating period.
			The short pump pulse further ensures the source is effectively instantaneous on the acoustic timescale, and the heat diffuses by at most a few hundred nanometres during the excitation.\cite{Kading1995,Kusnir2026}
			The resulting thermal expansion acts as a spatially harmonic force dipole at the surface,\cite{Rose1984} exciting an out-of-plane surface ripple whose full temporal evolution is recorded via diffraction of the continuous-wave probe laser.
			The frequency-domain amplitude of this signal, resolved as a function of propagation direction, constitutes the TGS maps shown in Figure~\ref{fig2}(a).	

	In the coordinate system introduced in Figure \ref{fig1}, the measured response to the TGS source should be described by the $G_{13}$ component of the frequency-domain Green's tensor {$G_{ij}(\mathbf{k}_\parallel,\omega)$} as introduced in Ref. \citenum{Every1997}, i.e., the component describing the out-of-plane displacement in response to an in-plane force.
	More details are given in the Methods section.
	Utilizing the partial-wave approach, $|G_{13}|$  maps were calculated for all experimentally studied cuts, with the result shown in Figure \ref{fig2}(b). 
	The agreement with the experimental maps is striking, regardless of which material and which cut is chosen.
	Even very minute variations in the calculated $|G_{13}|$ are matched in the {frequency-domain signal} amplitude $A$, and the shapes and relative intensities of all features are identical, up to some experimental scatter in $A$.
	
	The observed agreement confirms that, indeed, the modified TGS experiments are able to capture the additional features in the elastodynamic response, and to visualize the component of the frequency-domain Green's tensor in its whole complexity.
	This has two important implications: 
		First, the Green's tensor calculation can be used to explain several details in the experimentally obtained maps (which is an approach we adopt in the discussion below).
		And second, the TGS maps obtained with the modified setup bring rich information about the elastic behavior of the material, and thus, can be utilized for the determination of its anisotropic elastic coefficients $c_{ij}$. 
	
	For homogeneous, linearly elastic materials, the latter task is typically accomplished using an inverse approach.\cite{Landa2009a,Stoklasova2015}
	This means minimizing the mismatch between experimentally determined wavespeeds $v_n^{\rm exp}(\mathbf{m})$ in directions ${\mathbf m}$ and calculated values $v_n^{\rm calc}(\mathbf{m},c_{ij})$ for iteratively improved guesses of elastic constants,
	    \begin{equation}
	        \sum_n\left(v_n^{\rm exp}(\mathbf{m})-v_n^{\rm calc}(\mathbf{m},c_{ij})\right)^2\xrightarrow[{ c_{ij}}]{} \min.
	        \label{eq_inverse}
	    \end{equation}
	The richer the set of $v_n^{\rm exp}(\mathbf{m})$ (in terms of quantity as well as variety of wavetypes), the more complete information about the elastic constants is obtained.\cite{Stoklasova2015,Stoklasova2021}
	Several fine features in the TGS (and the $|G_{13}|$) maps, copy exactly the directional evolutions of wavespeeds of various acoustic modes, both surface-guided and limiting bulk waves, as seen from comparison of these evolutions (Figure \ref{fig2}(c)) with the maps.
	This means that extensive and rich sets of $v_n^{\rm exp}(\mathbf{m})$ can be extracted directly from measurements on a single free surface. 
	If the TGS experiments are performed on a principal cut, such as the $(1\,1\,0)$ cut for the Ni single crystal in the first column of Figure \ref{fig2}(a--c), some  points in the maps correspond to wavespeeds of bulk longitudinal and shear modes in principal directions that are related to $c_{ij}$ through simple algebraic equations, and thus, the elastic coefficients can be calculated without the inverse procedure.
	For general cuts and strongly anisotropic materials, in contrast, the use of the inverse procedure is inevitable, and it might be even difficult to identify the modes directly from the map, such as for the selected cuts in {Fe\textsubscript{3}Al}.
	Instead, approaches allowing for full visual comparison of these maps with their experimentally obtained counterparts would be more suitable, possibly replacing the summation in  (\ref{eq_inverse}) with integrals, and utilizing image-processing tools to measure the misfit between the calculation and the experiment. It is also worth noting that a single component of the frequency-domain Green's tensor appears to be fully sufficient for representing the anisotropic behavior of the studied material, as the wavespeeds of the SAWs/pSAWs as well as of most of the limiting bulk waves of various polarizations correspond to a visible contrast in the $|G_{13}|$ map. {Quantitatively, the match between contours in the TGS maps and corresponding contours in the $|G_{13}|$ plots, including contours marking wavespeeds of bulk waves, can be evaluated using the approach described in Methods. For the shown five maps, the root-mean-square (RMS) misfit between the experiment and the calculation does not reach above \SI{0.05}{\micro\metre\per\nano\second}, which is comparable to the accuracy of wavespeeds in time-domain methods \cite{sugawara2003,tachizaki2006,matsuda_time-resolved_2015,abbas_picosecond_2014,reverdy_elastic_2001,Stoklasova2015} discussed below.
} 
	
	The completeness of the information on elastic constants in the experimentally obtained maps turns TGS into a powerful tool in materials characterization.
	For single crystals, particularly those engineered for specific functionalities, the elastic anisotropy is often intrinsically linked to their caloric, magnetic, or electric properties.
	Such linking can be found in ferroelastic alloys with a strong elastocaloric effect for solid-state refrigeration\cite{manosa_materials_2017,xiao_giant_2022,hou_fatigue-resistant_2019,tusek_regenerative_2016,fahler_caloric_2018} and energy harvesting,\cite{bucsek_energy_2020,lindquist_mechanical_2023} or magnetoelastic materials for sensors and actuators.\cite{faran_ferromagnetic_2016,heczko_coupling_2022}
	For all these materials, the enhanced TGS characterization can bring additional insights in their functional properties and foster their further development.
	This  approach enables rich information to be obtained on a single cut of the crystal with arbitrary orientation.
	Moreover, the measurement spot is relatively local in terms of both its in-plane dimensions ($\sim$ 800 $\times$ \SI{800}{\micro\metre\squared}, can be reduced below $\sim$ 500 $\times$ \SI{500}{\micro\metre\squared} without any significant loss of information, as it still comprises several tens of fringes for the used wavelength) and its penetration depth (up to several wavelengths), which can be utilized for example for measurements on individual grains over surfaces of coarse-grained polycrystals, functionally graded materials with gradient composition, or individual components of macroscopic composites.
	In fact, the three cuts of {Fe\textsubscript{3}Al}\ presented here are a part of a single large polycrystalline piece of this material.

		For a finer spatial scanning, either the wavelength can be reduced or the number of fringes is lowered.
		The former leads to the need of detecting higher frequencies at the same quality to capture the ultra-transient features --- which may become technically difficult for frequencies reaching above \SI{1}{\giga\hertz}.
		The latter (i.e., by removing the number of fringes) is connected to lowering of the quality factor of the resulting spectrum (a very direct way to meet the Heisenberg principle, that is, that increasing the spatial precision is necessarily connected to lowering the frequency precision).
		This was presented in detail on the measurements by spatially resolved acoustic spectroscopy (SRAS),\cite{Smith2014} which was devised to scan SAW velocity across polycrystalline samples to evaluate the microstructure, crystallographic orientations, and, ideally, single-crystalline elasticity.\cite{Patel2017,Dryburgh2020,dryburgh_measurement_2022} 
		While the large number of SAW velocity measured in different crystallographic directions provides rich information, it still struggles to overcome the limitation of the single wave type to determine the full elastic tensor \cite{Stoklasova2015,dryburgh_measurement_2022}.

	\subsection*{Confirming the Ultra-Transient Gratings in the TGS Signal}
	
\begin{figure}[!t]
	        \centering
	        \includegraphics[width=\linewidth]{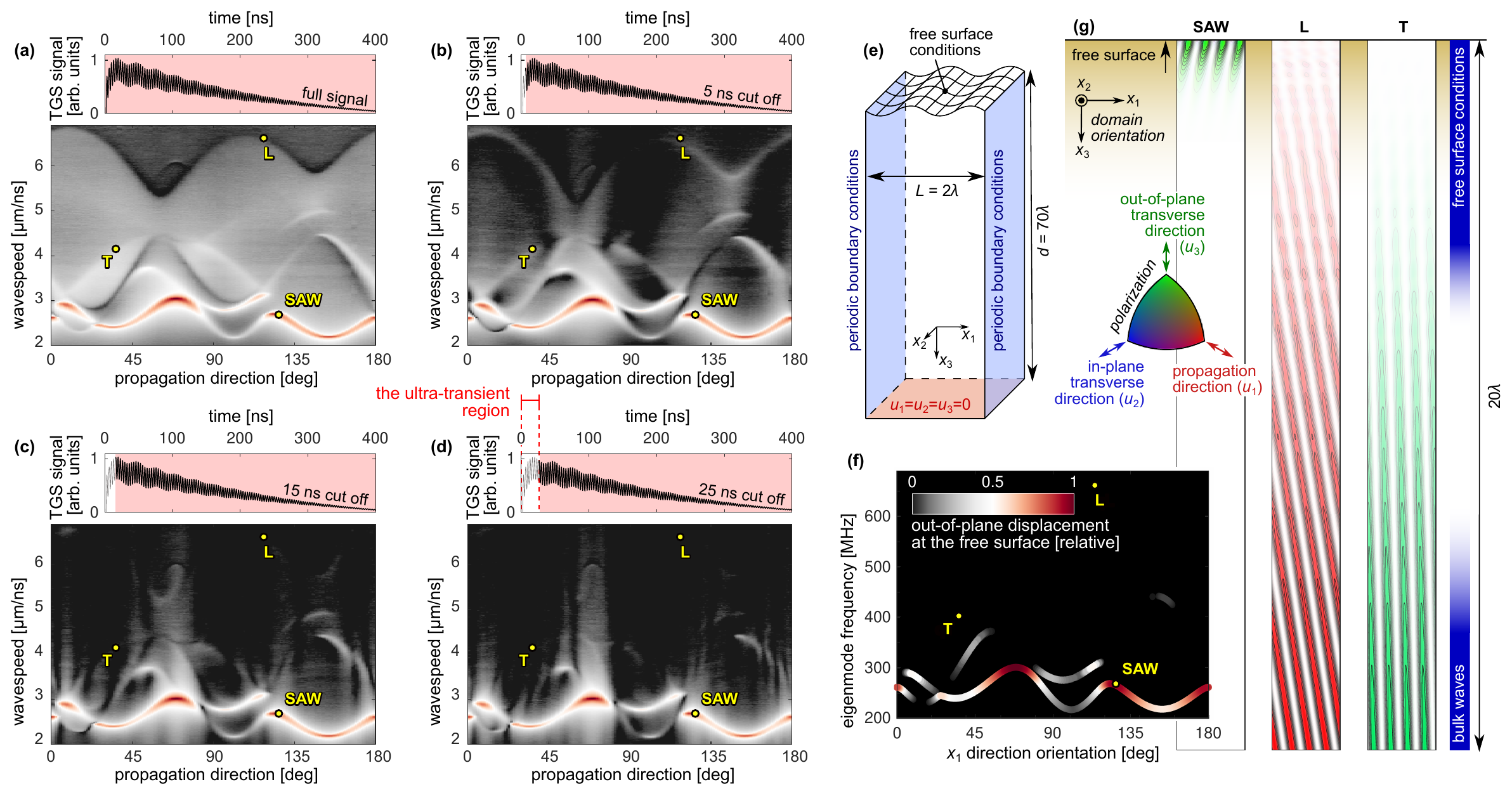}
	        \caption{
	            \textbf{The ultra-transient origin of detailed features in TGS maps.}\\
	            \textbf{(a-d)} Effect of time-domain signal truncation on the experimental TGS map for the {Fe\textsubscript{3}Al} \ (3\,1\,6) surface: Full TGS map (no truncation, (a)), followed by maps after omitting the first 5, 15, and \SI{25}{\nano\second}. After the ultra-transient (early-time) region is cut off (d), the map reduces to peaks indicating the (late-time) resonant response of the surface {(See the {\color{blue} Supplementary Discussion} for details on time-domain signal processing)}.
	            \textbf{(e-f)} Calculation of the resonant response of the surface using the Ritz--Rayleigh method: (e) the computational domain of dimensions $L\times{}d$ with boundary conditions for the displacement field ${\mathbf u}({\mathbf x})$; (f) the spectrum of eigenmodes of the domain, color-coded based on the out-of-plane component of the displacement field on the surface -- the spectrum perfectly matches the TGS map in (d); (g) modal shapes of the eigenmodes corresponding to the three selected points. The color in the maps indicates the polarization direction of the displacement, the intensity of the color represents an absolute value of the amplitude, ignoring the sign (two maxima per one wavelength $\lambda$). The labels L, T, and SAW in all subfigures refer to longitudinal (L), transverse (T), and surface (SAW) acoustic waves, respectively. }
	    	\label{fig3}
		\end{figure}

	To justify the above considerations, we need to prove that the additional features in TGS maps indeed originate from the immediate  response of the material to the thermoacoustic source.
	The visual agreement between the TGS maps and $|G_{13}|$ may suggest this is true, but direct evidence for that must be found by analyzing the time-domain signals.
	Such analysis is performed for one selected experiment (the approximately (3\,1\,6) cut of the {Fe\textsubscript{3}Al}\ crystal) in Figure \ref{fig3}:
	The subfigures (a)-(d) show how the spectrum of the angle-resolved signals (the TGS map) evolves as the earliest parts of the signals are progressively truncated.
	A representative time-domain signal corresponding to the first measured propagation direction is plotted above the maps.
	It is seen that the visible high-frequency oscillations last for hundreds of nanoseconds, corresponding to the lifetime of the transient grating arising due to the SAW/pSAW standing wave patterns. 
	Weak low-frequency oscillations appear in the signal, resulting from Scholte waves generated in the air above the free surface.\cite{glorieux2001}
	These will be excluded from further analysis.
	Note that the measurements were performed with the aforementioned acoustic wavelength $\lambda = \SI{10}{\micro\metre}$.
	
	In Figure \ref{fig3}(a), the full signal results in the full TGS map.
	Based on the curves in Figure \ref{fig2}(c), we can identify peaks or shoulders corresponding to specific wave propagating modes in this map (L -- longitudinal waves, T -- transverse (shear) waves, SAW -- surface acoustic waves) and select three points in which we will observe their evolution with truncation of the time-domain signal, each for one mode.
	Let us notice that, due to elastic anisotropy, the bulk modes are not purely longitudinal or purely shear waves, so the designations L and T refer to their dominant polarization orientation only.\cite{every_bulk_2013}
	Figure \ref{fig3}(b) then shows how the map changes with cutting off just \SI{5}{\nano\second} from the beginning of the signal. 
	The shoulders/edges marking the velocities of the longitudinal and shear modes become much less distinct, and even completely disappear for some directions of propagation.
	In the selected point L and T, some features are still visible, but their character changes from shoulders to very weak peaks that can be traced only because of the fine angular resolution in the experiment. 
	
	When \SI{15}{\nano\second} are cut from the beginning of the signal, any features indicating the longitudinal and shear modes in the selected point completely disappear (Figure \ref{fig3}(c)) and they disappear also for most of the propagation directions with cutting off the first \SI{25}{\nano\second}, Figure \ref{fig3}(d). 
	In other words, the information about all features in the map seen in Figure \ref{fig3}(a) but not in Figure \ref{fig3}(d) is contained in just less than \SI{25}{\nano\second} of the signal.
	And, as the most significant change is between subfigures (a) and (c), one can deduce that most of this information is included in the first \SI{15}{\nano\second} or less. 
	This means that the longitudinal modes (with wavespeed around \SI{6}{\micro\metre\per\nano\second}) travel less than \SI{90}{\micro\metre}, i.e., 9 wavelengths -- still, that is enough for such counter-propagating modes to form the ultra-transient grating, creating a feature in the diffracted signal.
	This means that the details in the TGS map indeed arise from phenomena that are transient with respect to the lifetime of the transient grating.
		
	After the ultra-transient region is removed from the time-domain signal, Figure \ref{fig3}(d), the TGS map becomes relatively simple. 
	It does not contain any shoulders, edges, or bands, just peaks that change their positions with the propagation direction; some of these peaks exist only in certain angular intervals, and there are discontinuous jumps in the positions of the peaks.
	The peaks are expected to represent local transient resonances of the surface, that is, standing waves.
	To confirm this assumption, we have performed an additional numerical calculation, summarized in Figures \ref{fig3}(e,f,g).
	We calculated vibrational eigenmodes of a domain shown in Figure \ref{fig3}(e) representing an elastic half-space with prescribed periodicity of the displacement field along the $x_1$ direction.
	The calculation was done using the Ritz--Rayleigh approach (see Methods) and resulted in a spectrum shown in Figure \ref{fig3}(f).
	In this plot, the curves show the evolution of the eigenmode frequencies as the domain orientation is rotated, and the color code represents the relative out-of-plane component of the surface displacement field for each eigenmode. 
	The calculation result agrees very well with the map in Figure \ref{fig3}(d). 
	This means that the vibrational response of the material beyond the ultra-transient region is indeed given by resonances of the surface in the form of standing wave patterns.
	From the three selected points in the map, the conditions for the resonant vibrations with non-zero out-of-plane amplitude are satisfied only for the SAWs.
	This is in a good agreement with the fact that the TGS experiments commonly capture the surface-guided modes, but not the limiting bulk acoustic waves.
	
	On the other hand, however, the points L and T also correspond to some eigenmodes of the domain.
	These are standing longitudinal and shear waves inside the bulk -- the displacement fields of these modes as they result from the Ritz--Rayleigh calculation are shown in Figure \ref{fig3}(g).
	The corresponding limiting bulk longitudinal and shear waves, albeit being solutions to the Christoffel's elastodynamic equation (Eq. (\ref{eq:Christoffel}) in the Methods section), do not satisfy the free-surface conditions; that is why the amplitude of the related eigenmode diminishes to zero when approaching the surface, and these modes thus do not result in any diffraction of the probe beam. 
	This holds true for all similar modes related to bulk acoustic waves with non-zero out-of-plane components of the polarization.
	In contrast, the eigenmodes representing standing Rayleigh-type and pseudo-Rayleigh surface waves (SAWs/pSAWs) are compliant with the free-surface conditions and have non-zero out-of-plane displacement field components.
	That is why they are observed in the late response to the source that persists in the measurement spot for several hundreds of nanoseconds.
	
	For completeness, let us mention that the requirement of a non-zero out-of-plane component of the displacement field applies also for the early-time response detected by UTGS.
	If any of the ultra-transient features have a strictly in-plane polarization, it remains invisible for the measurement.
	This is clearly seen for the high-symmetry (1\,1\,0) cut for the Ni single crystal in first column of Figure \ref{fig2}, where the UTGS contrast marking the fastest bulk shear wave completely disappears for propagation angles close to \ang{90}, where this mode becomes purely shear-horizontal.
	The absence of this type of waves in the out-of-plane response is identically reflected in the calculated $|G_{13}|$.

	\subsection*{Observed Acoustic Behavior of Anisotropic Free Surfaces}
Regarding the level of detail and completeness of the experimentally reconstructed $|G_{13}|$ maps, the UTGS results shown here provide more comprehensive directional and frequency-resolved information than the crystal-acoustics characterizations achieved through alternative experimental approaches, such as
        imaging of ballistic phonons, \cite{Wolfe_1998}
        picosecond ultrasonics, \cite{sugawara2003,tachizaki2006,matsuda_time-resolved_2015,abbas_picosecond_2014}
        or transmission acoustic microscopy.\cite{pluta2003}
The difference mainly stems from the use of point-like sources in these methods, so that the measured response is related to the time-domain Green's function, comprising wavefront arrival times.
As a consequence of this, fine two-dimensional spatial scanning is required to obtain the full direction-dependent information, and to allow the conversion of the resulting $G_{13}(\mathbf{x},t)$ to frequency-domain angle-resolved $G_{13}(\mathbf{k},\omega)$, where the level of detail in the $G_{13}(\mathbf{k},\omega)$ map is dictated by the spatial resolution of the scanning.
Note that while the $G_{13}(\mathbf{x},t)$  visualizes the anisotropic elastic behavior of the surface, the extraction of elastic properties of the material directly from it is difficult, or even impossible for low symmetry classes and general crystallographic cuts.
The reason is that the detected wavefront arrival times are given by the group velocity in the source-receiver direction, not the phase velocity, and
the wave-vector orientation in the corresponding waves is neither controlled nor known.\cite{audoin_measurement_1996} 
To resolve the latter issue, line-like sources can be used that enforce the wave propagation along the prescribed direction,\cite{reverdy_elastic_2001,Stoklasova2015} unless the energy flux is steered from this direction due to strong anisotropy,\cite{every_bulk_2013,Wolfe_1998, Musgrave_textbook} but the requirement on fine spatial scanning persists. 

In addition, the ballistic-phonon imaging, while providing visually compelling results, necessitates deep cryogenic temperatures and very short wavelengths to mitigate thermal scattering and phonon interactions.
And finally, although the pump-probe picosecond ultrasonics provides high sensitivity in a broad frequency bandwidth and allows for a detailed time-resolved study of the earliest parts of the signal,\cite{abbas_picosecond_2014,matsuda_time-resolved_2015}, it requires extensive temporal scanning in addition to the spatial one, and has, in general, limited ability to capture the late-time fields.

In principle, the UTGS approach overcomes these problems by the interference-pattern excitation which leads to strictly harmonic generation of coherent waves with well-defined wave-vector orientations (overcoming energy focusing and phonon scattering),\cite{Stoklasova2021} and a continuous-wave probe laser (possible due to lower frequencies involved). 
Besides UTGS, another opto-acoustic method offering angle-resolved analysis of the frequency-domain Green's tensor, specifically the $G_{33}$ component, is surface Brillouin spectroscopy. \cite{stoddart1998,every2016, Every2002}
However, the very low Brillouin-scattering efficiency means that even hours-long measurement of a single direction results in poor agreement between the measured spectrum and its calculated prediction beyond the SAW/pSAW peak.

\bigskip
The direct correspondence of UTGS to the $|G_{13}|$ component of the Green's tensor enables explaining theoretically the morphology of the experimentally observed TGS maps. 
	The so-called Lamb problem,\cite{Lamb1904} which means calculating the elastodynamic response of an anisotropic, linearly elastic half-space to a prescribed loading, can be solved semi-analytically using the concept of partial waves,\cite{Every1997} elaborated in more detail in the Methods section.
	In this approach, the displacement field in the elastic half-space for a given surface wavevector and frequency can always be decomposed into displacement fields of three independent solutions of Christoffel's elastodynamic equation (\ref{eq:Christoffel}).\cite{Musgrave_textbook}
	Each of these solutions is always either a planar harmonic bulk wave (the so-called homogeneous solution), or a surface-guided harmonic wave with its amplitude exponentially decreasing with the distance from the surface (the so-called evanescent solution).
	For a spatially periodic array of sources, the evanescent waves create transient gratings on the surface (as in Figure \ref{fig1}(f)), while the homogeneous waves create ultra-transient gratings (as in Figure \ref{fig1}(g)).
	When, for example, one of the partial waves converts from homogeneous into evanescent, the early-time response of the surface discontinuously changes, which results in a sharp contrast change in the TGS map.
	Indeed, as discussed below, except for the SAW/pSAW peaks, most of the sharp features seen in the $|G_{13}|$ plot, and, correspondingly, in the TGS maps, can be interpreted as changes in the types of the partial waves.
	
\begin{figure}[!t]
	       	\centering
	       	\includegraphics[width=\linewidth]{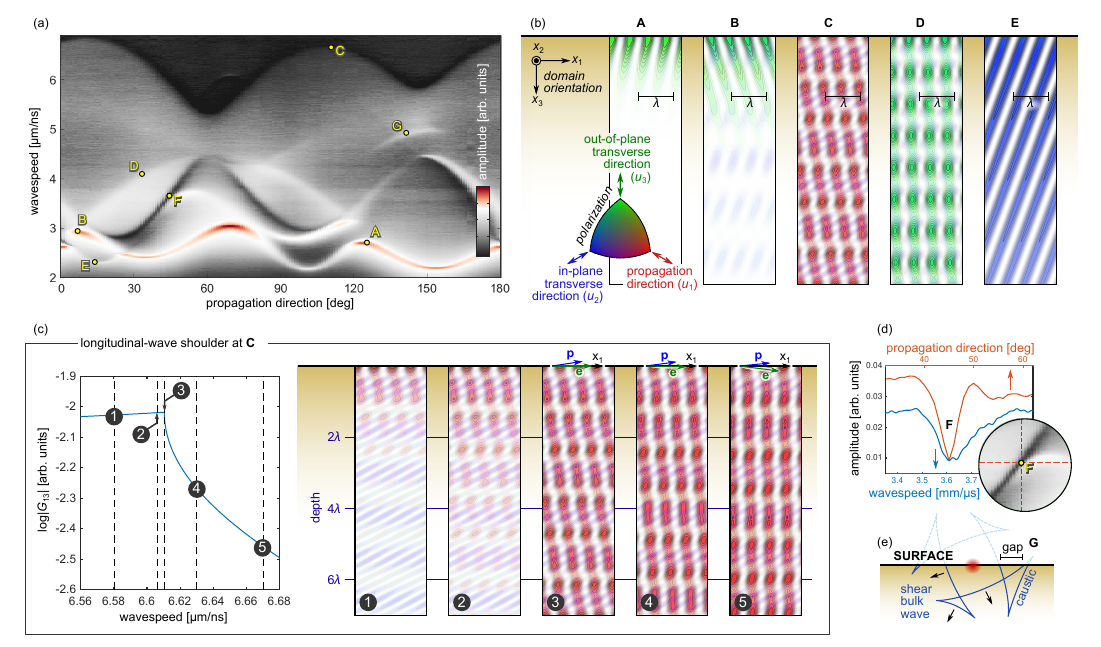}
	       	\caption{
	            \textbf{Detailed features in the TGS map explained by analysis of the Green's function.} \\
	            {\textbf{(a)}} TGS map for the {Fe\textsubscript{3}Al}\ (3\,1\,6) surface with selected features labeled by capital letters (A--G, points A, C, and D correspond to modes labeled SAW, L, and T in Figure \ref{fig3}, respectively);
	            {\textbf{(b)}} displacement fields for points A--E calculated as superpositions of three partial waves from the construction of the Green's function;
	            {\textbf{(c)}} the change in the displacement field (the evanescent-to-homogeneous conversion of one partial wave) responsible for the formation of a sharp shoulder (a cliff) at point C;
	            for stages (3)--(5), the arrows show the orientation of the energy flux direction ${\mathbf e}$ and a normal to the longitudinal phase fronts ${\mathbf p}$ with respect to the propagation direction $x_1$.
	            {\textbf{(d-e)}} narrow-band features observed in the map: the acoustic sink (in (d), cuts w.r.t. both coordinates are shown) and the caustics (in (e), schematic explanation of the creation of a disconnected segment).
	             }
	                   	\label{fig4}
	    \end{figure}

	\medskip
	The partial wave approach enables constructing the displacement fields corresponding to {individual points in the $|G_{13}|$ plot,} so that one can directly calculate which orientations and types of elastic strain field are carried by the individual features in the TGS map.
	An example of such analysis is given in Figure \ref{fig4}, using, similarly as for Figure \ref{fig3}, the (3\,1\,6) cut for the {Fe\textsubscript{3}Al}\ crystal.
	In Figure \ref{fig4}(a), the TGS map for this cut is shown, with selected points of interest (capital letters A-G) in which we utilize the calculated counterpart of the TGS map to interpret the observation.  
	
	The points A--E correspond to sharp edges and peaks in the map. 
	According to Figure \ref{fig2}(c), these sharp features mark the velocities of SAWs, pSAWs and limiting bulk acoustic waves, respectively.
	The partial wave approach enables us to confirm this by explicitly calculating the displacement fields, as shown in Figure \ref{fig4}(b).
	The point A is selected identically to the point labeled SAW in Figure \ref{fig3}.
	It corresponds to a Rayleigh-type SAW, which is composed of three evanescent solutions of the Christoffel equation (\ref{eq:Christoffel}).
	As this wave is included both in the early and late responses, the displacement field for this wave calculated from the partial waves is identical to that obtained from the Ritz--Rayleigh calculation of the eigenmodes.
	A similar situation appears for the point B, which corresponds to a pSAW mode.
	The displacement field has two evanescent components with dominantly in-plane polarizations, but includes also one homogeneous, out-of-plane polarized wave responsible for the sharp peak in the TGS map. 
	Again, a resonant mode approximating closely the pSAW standing pattern is included in the eigenproblem solution and has a strong out-of-plane component, and thus the peak at point B persists in the spectrum even after the ultra-transient region is cut off (Figure \ref{fig3}(d)).
	
	The point C corresponds to the point labeled L in Figure \ref{fig3}, as it marks a bulk longitudinal mode, absent in the late-field response.
	The displacement field for this point in Figure \ref{fig4}(b) reveals that this mode is, indeed, dominantly polarized in the longitudinal direction (along the propagation direction). 
	However, to comply with the prescribed harmonic loading at the surface, the dominant homogeneous longitudinal wave is coupled with two other homogeneous partial waves that disturb the planar wavefronts of the longitudinal mode and result in a more complex displacement field.
	The wavespeed associated with this displacement field corresponds to the dominant longitudinal wave, and the other two partial waves can be understood as resulting from the interaction of this dominant wave with the surface.
	The TGS and $|G_{13}|$ maps in this point exhibit a cliff (a sharp shoulder for the given direction of propagation), which results from conversion of one of the partial waves as we discuss below in Figure \ref{fig4}(c).
	A very similar situation occurs for point D (identical to the point labeled T in Figure \ref{fig3}), which corresponds to the velocity of a limiting bulk shear wave with out-of-plane polarization.
	Again, the displacement field is a superposition of three partial waves, among which the out-of-plane polarized one is the strongest, and the cliff associated with this point arises by a similar mechanism as for point C.
	
	A weak but still visible contrast is seen at point E.
	The velocity corresponding to this point equals the velocity of an in-plane polarized bulk shear wave that, in principle, should not be detectable by the probe beam diffraction.
	Accordingly, this point does not correspond to any detectable eigenmode in Figure \ref{fig3}(f).
	However, the early field response analysis reveals that the contrast arises due to a transition of a weak out-of-plane polarized homogeneous wave from evanescent to homogeneous;
	at the surface this wave effectively couples with other two partial evanescent waves, resulting in a dominantly in-plane shear polarization, as seen in Figure \ref{fig4}(b).
	
	Figure \ref{fig4}(c) gives an explanation of how the sharp cliffs marking the limiting bulk wave velocities are created.
	It shows the evolution of the displacement field in the vicinity of the point C as the wavespeed coordinate increases.
	For wavespeeds lower than the velocity of longitudinal limiting bulk waves (stages (1) and (2) in Figure \ref{fig4}(c)) at least one of the partial waves is evanescent, and the energy of the waves remains dominantly bound to the surface.
	As the wavespeed coordinate increases, the evanescent character of the displacement field becomes weaker, and, finally, the given partial wave becomes homogeneous at the wavespeed equal to the velocity of the limiting longitudinal bulk wave (stage (3) in Figure \ref{fig4}(c)).
	At this point, the energy flux carried by the wave points exactly along the surface, as required by the definition of the limiting bulk wave.
	With further increasing the wavespeed coordinate, however, the orientation of the longitudinal wave fronts starts rotating (stages (4) and (5) in Figure \ref{fig4}(c)), and so does the orientation of the energy flux, which means that the acoustic energy becomes more and more steered away from the surface.
	In other words, the cliff indeed represents a conversion from an evanescent solution, in which the energy is bound to the surface (as in Fig.\ref{fig1}(f)), to a homogeneous solution that drives the energy below the surface (as in Fig.\ref{fig1}(g)).
	In the experimental map, the cliff from the early field response can be partially smoothed or overlapped with a small peak formed by the surface skimming longitudinal waves\cite{Stoklasova2021} that travel at wavespeeds bounded from above by the velocity of the limiting bulk longitudinal waves.

	\medskip
	When two cliffs form close to each other, they can encapsulate a narrow band as the one seen in the point F.
	In Figure \ref{fig4}(d) cross-sections of the experimental TGS maps in the surrounding of the point F are shown, documenting the sharp drop in the amplitude along both the wavespeed and the angle coordinates.
	The free surface in this area acts as a highly selective filter that transmits all waves except those with specific velocity.
	On the other hand, the band between two cliffs cannot be understood as a band gap similar to those appearing in periodic media, where the energy at given frequency is trapped in a local resonance of the structure.
	Here, instead, the energy at a specific frequency is steered from the surface, as, in a narrow frequency interval, the character of one partial wave changes from evanescent to homogeneous and back.
	Hence, the feature in point F can be seen as something that we term an acoustic sink, where only waves at very specific wavespeeds are allowed to penetrate into the material.
	
	An inverse feature to the acoustic sinks can be seen in point G, where the amplitude of the TGS map locally increases and creates an object disconnected from other edges or peaks in the map.
	Similar features appear in all low-symmetry cuts, and are associated with disconnected segments of bulk shear waves.
	These arise from the complex, cuspidal shapes of bulk shear wave slowness surfaces (caustics) intersecting the free surface (Figure \ref{fig4}(e)).
	Such intersections can enable bulk wave energy to briefly travel along the surface or provide pathways for surface energy to couple into angle-focused bulk beams.
	The edges of these caustics are typically directions of strong energy focusing,\cite{Wolfe_1998,Musgrave_textbook} likely contributing to the relative prominence of these features in the maps. 
	In terms of partial waves, these objects are formed by a mechanism very similar to the acoustic sinks, but in this case the evanescent solution locally changes into a homogeneous solution that carries energy along the surface, rather than steering it into the material.

\section*{Discussion}
Due to the emergence of the ultra-transient gratings, the transient grating spectroscopy provides an experimental capability to visualize the complete surface acoustic response in elastically anisotropic materials, as we have demonstrated here on benchmark examples of two prototypical cubic single crystals.
This goes far beyond conventional surface acoustic wave characterization that TGS has been designed for, bringing direct and detailed insight into the acoustics of the studied crystals.
By capturing and analyzing the rich information encoded within the initial nanoseconds of the elastodynamic response (which we call the ultra-transient region), the UTGS approach generates experimental maps that strikingly replicate the corresponding $|G_{13}|$ component of the theoretical elastodynamic Green’s function in all its complexity.

The finer structure of the frequency-domain Green's function corresponds to the behavior of partial waves the elastodynamic response of the surface is composed of, at a given frequency and wavevector.
With conversions between homogeneous and evanescent partial waves, a sharp change in the ultra-transient part of the signal arises, leading to a detectable contrast variation in the TGS map.
Not only that these contrast variations enable determining the velocities of bulk elastic modes, which is beneficial for inverse determination of elastic constants.
In addition, they directly evidence acoustic energy steering from the surface into the bulk by homogeneous partial waves, or conserving it along the surface by evanescent partial waves---phenomena that arise from superpositions of wave fields of arrays of coherent acoustic sources on the surface.

The all-optical contactless nature of the UTGS analysis predetermines the presented approach to be used under various extreme conditions (low and high temperatures, strong magnetic fields, irradiation), enabling direct in-situ observation of the frequency-domain $|G_{13}|$ components of materials under these conditions with time or with additional external stimuli.
This opens a pathway for studying the relationship between elastic behavior and performance in functional materials.

\bigskip

\section*{Methods} \label{sect:Methods}
	\subsection*{Samples}
        The single-crystalline samples of nickel were acquired from MaTecK (Material, Technologie \& Kristalle GmbH), with the accuracy of the crystallographic orientation of $<\ang{0.1}$.
        The considered density was $\rho = \SI{8.9}{\gram \per \centi\metre\cubed}$, and the elasticity of nickel was taken as $c_{11} = \SI{251.9}{\giga\pascal}$, $c_{12} = \SI{157.7}{\giga\pascal}$, and $c_{44} = \SI{124.3}{\giga\pascal}$.\cite{Stoklasova2021}
        {
			The absorption coefficient at the pump wavelength was $\zeta = \SI{0.139}{\per \nano\metre}$, and the thermal diffusivity was $\alpha = \SI{23.2}{\square \milli \metre \per \second}$.\cite{werner2009,emsley1998}
		}

        The {Fe\textsubscript{3}Al}\ samples were cut from an intermetallic alloy (exact composition Fe--28 at.\%Al--4 at.\% Cr) manufactured by Bridgman growth at the Institute of Physics, Czech Academy of Sciences.
        The density was $\rho = \SI{6.62}{\gram \per \centi\metre\cubed}$, and elasticity $c_{11} = \SI{179.5}{\giga\pascal}$ , $c_{12} = \SI{119.6}{\giga\pascal}$, and $c_{44} = \SI{129.9}{\giga\pascal}$.\cite{Kusnir2023}
        The precise orientations of the surfaces were determined by X-ray diffraction (Laue method), as summarized in {\color{blue} Supplementary Table S1}.
        {
			The absorption coefficient at the pump wavelength was estimated to be on the same order as for Ni ($\zeta \approx \SI{0.1}{\per \nano\metre}$), and the thermal diffusivity was taken as\cite{rudajeva1997} $\alpha = \SI{3.3}{\square \milli \metre \per \second}$---i.e., nearly an order of magnitude lower than that of Ni.
		}
     
		All measured surfaces were mechanically and chemically polished to obtain a good reflectivity (mirror-like finish) for the probe laser.

	\subsection*{Transient grating spectroscopy}
    	The measurements were performed using the optical path outlined in Figure \ref{fig1}(a), which is a classical arrangement for TGS.\cite{Maznev1998,Stoklasova2021}
        The pump pattern was created by interference of strongly elliptical laser beams 
            (approximately 0.8 $\times$ \SI{0.1}{\milli\metre\squared} in cross-section)
        in the near infrared range 
            (\SI{1064}{\nano\metre}, pulse energy \SI{110}{\micro\joule}, pulse duration \SI{0.53}{\nano\second} (FWHM), repetition rate \SI{1}{\kilo\hertz}).
        The nominal acoustic wavelength in the reported measurements was \SI{10}{\micro\metre}, with the exact wavelength calibrated using a known material -- for the reported measurement, the calibrated value was $10.064 \pm \SI{0.005}{\micro\metre}$.
        The pulse energy was lowered in the optical path and tuned to be slightly below the ablation limit of the sample (typically below \SI{40}{\micro\joule} on the sample surface), so that no visible trace was left after the measurement. 
		The heat diffusion length (approximated as $L = \sqrt{\alpha t}$) was well below the acoustic wavelength (\SI{10}{\micro\metre}): around \SI{0.48}{\micro\metre} for Ni, and \SI{0.18}{\micro\metre} for {Fe\textsubscript{3}Al}{} at \SI{10}{\nano\second} after the laser pulse.

    	The probe-beam pattern, created by a continuous-wave green laser
            (\SI{532}{\nano\metre}, nominal power \SI{100}{\milli\watt}, reduced to \SI{30}{\milli\watt} at the sample surface), 
        was aligned with the same optical axis to further enhance the $k$-vector selectivity of the method.\cite{maznev_surface_1999}
        To increase the overall sensitivity, the system utilizes the heterodyne detection approach.\cite{Kading1995,Maznev1998,Johnson2012,Verstraeten2015}
        The heterodyne phase was adjusted for optimal sensitivity to the out-of-plane displacements using a phase retarder.
			Note that the heterodyne phase varied by a few degrees during the signal averaging; however, this had no significant effect as the response was flat around the optimal phase. \cite{Dennett2018}

        The diffracted probe beam was collected by a Si PIN photodiode (with maximal continuous input power of \SI{50}{\milli\watt} and photosensitivity of \SI{0.38}{\ampere\per\watt} for the probe-laser wavelength), and the signal was amplified with the gain of \SI{60}{\decibel} in the frequency bandwidth of \SI{10}{\kilo\hertz} to \SI{1}{\giga\hertz}. 
		The signal was digitized with a sampling rate of \SI{5}{\giga\hertz} and 10-bit resolution.
			The whole signal-acquisition system had a flat frequency response in the range \SIrange{10}{700}{\mega\hertz}.

    	To perform the angular-dispersion measurements, the sample was rotated with respect to the optical axis. 
        At each measured angle, fifty thousand time-domain signals were averaged.        
        {The averaging led to longer acquisition times: the repetition rate of the excitation laser was \SI{1}{\kilo\hertz}, but the effective rate was lowered by the signal acquisition to around \SI{600}{\hertz}.
			Additional time was needed for the sample rotation and for the data transfer and storage; thus, the total time for one angle was around \SI{1.5}{\minute}, and  a full TGS map took approximately \SI{5}{\hour}. 	Further details on the signal acquisition and processing---including the subtraction of background (parasitic) signals, signal cutting/windowing, the influence of the laser-pulse duration on the frequency response, the effect of angular resolution and the number of averages on the resulting maps, and evidence of reproducibility across different experimental setups---are provided in the {\color{blue} Supplementary Discussion}.}

    \subsection*{Calculation of the Green's function {and its interpretation with respect to the experiment}}
        The {frequency-domain} surface dynamics calculation was based on the approach by Every et al.,\cite{Every1997,Every2015} which can be summarized as follows:
        
        For the given tensor of elastic constants $C_{ijkl}$, material density $\rho$, angular frequency $\omega$, and surface wavevector $\mathbf{k}_\parallel = (k_1,k_2)$, the Christoffel equation
    		\begin{equation}
    			\left( C_{ijkl} k_j k_l - \rho \omega^2 \delta_{ik} \right) U_k = 0 \label{eq:Christoffel}
    		\end{equation} 
    	gives six solutions for the out-of-plane component of the wavevector $k_3^{(n)}$, $n=1,\ldots,6$, with corresponding eigenvectors $\bf{U}^{(n)}$. 
		The waves described by the resulting three-dimensional wavevectors $\bf{k}^{(n)}$ and polarizations $\bf{U}^{(n)}$ are the partial waves.
    	Three of these waves represent either the evanescent waves (complex solutions in $k_3$ for which the displacement field decreases exponentially along the $x_3$ direction) or the outgoing waves (real solutions of $k_3$, for which the group velocity is directed into the solid).
    	
    	The displacement resulting from a spatially and temporally harmonic surface force $F_j$ (acting in the $x_j$-direction and defining $\mathbf{k}_\parallel$ and $\omega$) is then formed as a superposition of these three phase-matched partial waves,
            \begin{equation}
                u_i^{(j)} (\mathbf{k}_\parallel, \omega; x_3) = \sum_{n=1}^3 A_j^{(n)} U_i^{(n)} \exp (\mathrm{i} k_3^{(n)} x_3) , \quad (i,j = 1,2,3). 
            \end{equation}
        The partial-wave amplitudes $A_j^{(n)}$ stem from the stress-strain relationship and the free-surface boundary condition as
            \begin{equation}
                A_j^{(n)} = \frac{\rm i}{\omega} \left(\mathbf{B}^{-1} \right)_j^{(n)},  \quad (j,n = 1,2,3).
            \end{equation}
        where the boundary condition matrix can be written as
    		\begin{equation}
    			B_j^{(n)} = C_{3jkl}U_k^{(n)} \frac{k_l^{(n)}}{\omega}, \quad (l,n = 1,2,3).
    		\end{equation}
        The surface Green's function components are then obtained as the displacement amplitude at $x_3 = 0$, which reads,
        		\begin{equation}
        			G_{ij}(\mathbf{k}_\parallel, \omega) =
                        \sum_{n=1}^3 A_i^{(n)} U_j^{(n)}.
        		\end{equation}

			To connect Green's function and the TGS measurement, we followed the model introduced by Maznev et al.\cite{Maznev1999} that rests on approximating the thermoelastic source as an instantaneous and surface-bound heating.
			Given the absorption coefficients specified above, the optical penetration depth  at the pump wavelength is below \SI{10}{\nano\metre} for the studied samples, and we can safely assume that all heat is generated in a thin layer near the surface.
			Under this approximation, the thermoelastic source is composed of horizontal forces arising in response to the thermal expansion of the surface\cite{Rose1984} and lying along the direction perpendicular to the fringes ($x_1$). 

			The resulting surface ripple is detected by the probe-beam diffraction, where the diffracted signal intensity is proportional to the amplitude of the slope of the surface profile, and, given the harmonic nature of the ripple, also to the out-of-plane ($x_3$) surface displacement amplitude. As a result, the Fourier transform of the TGS signal in the frequency domain is directly proportional to the magnitude of the $G_{13}$ component of the Green's tensor, which describes the $x_3$ (out-of-plane) displacement at the surface in response to an $x_1$-oriented (horizontal) force. In particular, the standing-wave response to the TGS source is composed of $G_{13}(\mathbf{k}_\parallel, \omega)$ and $G_{13}(-\mathbf{k}_\parallel, \omega)$, that are not necessarily equal for anisotropic materials and general crystallographic cuts. The superposition magnitude is
					\begin{equation}
						|G_{13}|_{\rm eff.} =  |G_{13}(\mathbf{k}_\parallel, \omega) + G_{13}(-\mathbf{k}_\parallel, \omega)|.
					\end{equation}
			which is the quantity plotted in Figure \ref{fig2}(b); for simplicity, we denote it as $|G_{13}|$ in the main text and the figures, without the subscript.

			Finally, let us point out that the Green's function calculation in Refs.\citenum{Every1997,Every2015} assumes a harmonic source in both space and time. 
            Spatially, the central part of the interference pattern provides a very good approximation of the required harmonic source. Temporally, the source is a pulse, broadband in the frequency domain, which means the Fourier transform of the time domain signal provides a good approximation of sweeping through the range of frequencies in the Green's function calculation, although the non-flat frequency characteristic of the real source leads to a decline in excitation amplitude with increasing frequency. The used laser-pulse duration of \SI{0.5}{\nano\second} is relatively short compared to the acoustic timescales of SAWs (with frequency below \SI{300}{\mega\hertz}, meaning period above \SI{3.3}{\nano\second}). For the higher frequencies, the pulse duration is not negligible, but still short enough to allow for significant excitation of the acoustic waves in the studied frequency range (up to \SI{1}{\giga\hertz}, meaning period of \SI{1}{\nano\second}).       
            While a shorter pulse would improve the high-frequency response, the current pulse duration is sufficient for the frequencies of interest in this work and does not compromise the equivalence between the TGS maps and the Green's function, as evidenced by the agreement reported in Figure \ref{fig2}. More details on the effect of the pulse duration are reported in the {\color{blue} Supplementary Discussion.}

	\subsection*{Assessing the quantitative agreement between experiment and calculation}
	To quantify the agreement between the calculated wavespeeds or features in $|G_{13}|$ plots, and experimental TGS maps, we used a sequential three-step procedure described below. In the first step, we compared the TGS map contrasts marking wavespeeds of SAW/pSAW waves and longitudinal waves to the calculated wavespeeds. Here, a very good match was expected a priori, because the SAW/pSAW data and partially also the longitudinal wavespeed data obtained on the same materials were used in our previous works\cite{Stoklasova2021,Kusnir2023} to determine the elastic constants of nickel and {Fe\textsubscript{3}Al}{} through minimization (\ref{eq_inverse}). In the second step, we extracted additional features from the TGS maps that marked wavespeeds of limiting bulk shear waves. Regions where the extracted contours were intersecting or running close to each other were excluded from the quantitative analysis, to avoid ambiguity in pairing between individual curves in the experiment and the calculation. The extracted data were then compared with calculated wavespeeds. In the final step, we did the comparison for sharp features seen in both the experimental TGS maps and the $|G_{13}|$ plots, but did not have direct counterparts in the calculated wavespeed data.
	
	In other words, the first step documented the correctness and accuracy of the elastic constants used for the calculations, the second step documented the accuracy with which the shear bulk wave data can be extracted from the TGS maps, and the third step illustrated the ability of the UTGS mechanism to reproduce features in the Green's function beyond the wavespeed data. Although these last features cannot be involved in the minimization (\ref{eq_inverse}), their goodness-of-fit can serve as the independent verification of correctly determined elastic constants.

	In {\color{blue} Supplementary Figure S2}, the application of these three steps is shown for the approximately $(1\,2\,7)-$oriented cut of  {Fe\textsubscript{3}Al}{}. We will use this cut to illustrate the procedure---the results for all other cuts and both materials were qualitatively and quantitatively very similar. The dots in {\color{blue} Supplementary Figure S2} show locations of peaks and edges, extracted from the maps using numerical image-processing tools: the former were detected utilizing Lorentzian mask fitting, the latter by using the Canny (maximal slope) edge-detection algorithm\cite{Canny1986}. For the edge detection, the data was pre-smoothed using a Savitzky-Golay filter with the window width of 0.2 $\mu$m$\cdot$ns$^{-1}$. Especially the edge detection was done semi-manually, with selecting the approximate locations first from visual assessment of the maps, and then applying the pre-smoothing and numerical detection of the maximum slope in the given intervals.
	
	Within the first step, the dominant peak locations (SAW/pSAW data) and upper edges of the maps (longitudinal wavespeeds) were extracted, as shown in {\color{blue} Supplementary Figure S2} in light green color.  It is seen that the calculated wavespeeds of the corresponding modes matched these data very well. For the SAW/pSAW wavespeeds, the maximum misfit over the whole dataset was \SI{0.058}{\micro\metre\per\nano\second}, and the RMS error was \SI{0.022}{\micro\metre\per\nano\second}. For the longitudinal wavespeeds, the maximum misfit was \SI{0.101}{\micro\metre\per\nano\second} (the maximal misfits were attained at minimal wavespeeds, where the edges were less pronounced, and thus, their localization more difficult), and the RMS error was \SI{0.044}{\micro\metre\per\nano\second}.
	
	Within the second step (light blue color in {\color{blue} Supplementary Figure S2}), additional features were extracted from the TGS maps using the same numerical tools; most of these features matched some of the calculated bulk shear wavespeed curves. The maximum misfit was \SI{0.074}{\micro\metre\per\nano\second}, and the RMS error of the fit was \SI{0.038}{\micro\metre\per\nano\second}, which means the same accuracy as for the SAW/pSAW and longitudinal data that were used for the determination of the elastic constants.
	
	Finally, within the  third step (dark red color in {\color{blue} Supplementary Figure S2}), features from the $|G_{13}|$ plot that are not directly connected to wavespeeds of limiting bulk waves were compared with the experiment. In the given cut, these features are mainly the prominent upper edge of an acoustic sink (see the main text) between 90$^\circ$ and 110$^\circ$, and the narrow cloudy contrast occurring at wavespeed faster than most of the bulk shear waves between approximately 40$^\circ$ and 70$^\circ$. Locations of these features match the experimental data with the maximum misfit of \SI{0.122}{\micro\metre\per\nano\second}, and RMS error of \SI{0.048}{\micro\metre\per\nano\second}.

    \subsection*{Ritz--Rayleigh calculation method}
        The eigenmodes of the domain representing the free surface (Figure \ref{fig3}(e)) were calculated using the  Ritz--Rayleigh method for SAWs in an elastically anisotropic half-space, introduced and described in detail in Ref. \citenum{Stoklasova2015}. 
        The domain depth was set as $d=70\lambda$ because the penetration depth of some of the evanescent modes increases with increasing strength of elastic anisotropy ($d=20\lambda$ would be fully sufficient for the Ni crystal).\cite{Kusnir2023}
        This depth enabled solutions representing both the near-surface behavior and the bulk behavior.
        Because of its assumed homogeneity in $x_2$, the displacement field $u_i$ ($i=1,\ldots,3$) was discretized into a functional basis as
    	        \begin{equation}
    	        	u_i (x_1, x_3) =
    	        	\sum_{k=0}^{N-1} \alpha_{k,i} P_{k}\left(2\hat{x}_3-1\right) \exp(2 \pi \mathrm{i} \hat{x}_{1}) 
    	        	,
    	        	\label{eq:RRbasis_SAW}
    	        \end{equation}
            with $\hat{x}_3=x_3/d\in[0,1]$ and $\hat{x}_1 = x_{1}/\lambda \in[0,1]$.
            Here, $P_{k}(x)$ denotes a normalized Legendre polynomial of the $k$-th order, and $\alpha_{k,i}$ are coefficients of the functional basis.
	          Legendre polynomials were used up to $N=180$ in the discretization.
	          The chosen functional basis inherently satisfied the periodic boundary condition.
	          The bottom-plane constraint was imposed using a null-space method.\cite{Stoklasova2015,Deng2022}
	          
	          The eigenmodes and eigenfrequencies of the domain were then sought using the Hamilton principle for the discretized displacement field, which has a form of the eigenvalue problem in a matrix form
	        \begin{equation}
	        	0=(\omega^{2}\mathbf{M}-\mathbf{K})\bm{\upalpha},
	        	\label{eq:RRfinal}
	        \end{equation}
	        where $\bm{\upalpha}$ is the vector of the coefficients from the discretization.
	         Dimensions of matrices $\bf M$ and $\bf K$ are given by the size of the computational basis as $(3N-1)^2$.
            More details and an explicit form of these matrices can be found in previous works of the authors.\cite{Landa2009a,Sedlak_21konst,Stoklasova2015,Grabec2024_NiTi}
            The resulting eigenvalues represent the square of the frequency $\omega$ of the corresponding mode.
            To describe the character of the modes, the displacement field can be reconstructed using its eigenvector $\bm{\upalpha}_k$ (as used for Figure \ref{fig3}(g)).
            However, for various characterizations of the properties of the mode, eigenvectors can be used without the need to reconstruct the modal shape -- e.g. to trace the angular-dispersion curves for individual modes, the similarity of modal shapes can be evaluated quantitatively by the absolute value of the dot product of their eigenvectors.\cite{Grabec2020}

\section*{Acknowledgments}
The authors acknowledge Dr. Miroslav Frost (Czech Acad. Sci.) for help with open data management.

\section*{Funding}
This work has been financially supported by Czech Science Foundation [Project No. 22-13462S] and by the Operational Programme Johannes Amos Comenius of the Ministry of Education, Youth and Sport of the Czech Republic, within the frame of project Ferroic Multifunctionalities (FerrMion) [project No. CZ.02.01.01/00/22\_008/0004591], co-funded by the European Union.

\section*{Author contributions statement}
T.G., P.Sed. and H.S. conceptualized the work and proposed the theoretical explanation for the observations. P. Sto. and D.M. designed, developed and constructed the dedicated modifications of the TGS setup, and M.\v{S}. developed tools for experiment automation and data acquisition. K.R. and J.K. carried out the experiments and performed post-processing of the experimental data. T.G. and J.K. performed numerical simulations. The initial draft of the text was written by T.G. and H.S., and reviewed, commented on, and edited by all other authors (P.Sto., K.R., J.K., D.M., M.\v{S}. and P.Sed.).)

\section*{Competing interests}
The authors declare no competing interests.

\section*{Data availability statement}

The data generated in this study have been deposited in the Zenodo repository under accession code https://doi.org/10.5281/zenodo.16875359. The source data are identical to the data available under this accession code.

\end{document}